\documentclass[12pt, a4paper, reqno]{amsart}
\usepackage{mathrsfs}
\usepackage{bbm}
\usepackage{amsmath,amscd,amssymb,latexsym}
\usepackage[all]{xy}
\usepackage{color}
\usepackage[section]{placeins}
\input xypic

\evensidemargin=\oddsidemargin
\DeclareMathVersion{can}
\DeclareMathAlphabet{\can}{OT1}{cmss}{m}{n}
\newtheorem{thm}{Theorem}[section]
\newtheorem{cor}[thm]{Corollary}
\newtheorem{lem}[thm]{Lemma}
\newtheorem{prop}[thm]{Proposition}
\newtheorem{rem}[thm]{Remark}
\newtheorem{exa}[thm]{Example}
\theoremstyle{definition}
\newtheorem{defn}[thm]{Definition}
\theoremstyle{fact}

\theoremstyle{conjecture}

\numberwithin{equation}{section}

\newcommand{\ord}{\operatorname{ord}}

\newcommand{\Tr}{\operatorname{Tr}}

\begin{document}

\title[]{Binary irreducible quasi-cyclic parity-check subcodes of Goppa codes and extended Goppa codes}
\author [Li] {Xia Li}
\address{\rm Department of Mathematics, Nanjing University of Aeronautics and Astronautics,
Nanjing,  211100, P. R. China}
\email{lixia4675601@163.com}

\author[Yue]{Qin Yue}
\address{\rm Department of Mathematics, Nanjing University of Aeronautics and Astronautics,
Nanjing,  211100, P. R. China}
\email{yueqin@nuaa.edu.cn}
\author[Huang] {Daitao Huang}
\address{\rm
College of Computer Science and Technology, Nanjing University of Aeronautics and Astronautics,
Nanjing,  211100, P. R. China}
\email{dthuang666@163.com}

\thanks{The paper was supported by National Natural Science Foundation of China (No.
 61772015),  the Postgraduate Research $\&$  Practice Innovation Program of Jiangsu Province (No. SJKY$19_{-}0167$).}
\begin{abstract}
Goppa codes are particularly appealing for cryptographic applications. Every improvement of our knowledge of Goppa codes is of particular interest.
In this paper,  we present a sufficient and necessary condition for an irreducible monic polynomial $g(x)$ of degree $r$ over $\mathbb{F}_{q}$ satisfying $\gamma g(x)=(x+d)^rg({A}(x))$, where $q=2^n$, $A=\left(\begin{array}{cc} a&b\\1&d\end{array}\right)\in PGL_2(\Bbb F_{q})$, $\mathrm{ord}(A)$ is a prime, $g(a)\ne 0$, and $0\ne \gamma\in \Bbb F_q$.   And we give a complete characterization of  irreducible polynomials $g(x)$ of degree $2s$ or $3s$ as above, where $s$ is a positive integer. Moreover, we construct some binary irreducible quasi-cyclic parity-check subcodes of Goppa codes and extended Goppa codes.
\end{abstract}

\maketitle

\section{Introduction}
Goppa codes are particularly appealing for cryptographic applications. McEliece
was the first to exploit the potential of Goppa codes for the development of a secure
cryptosystem \cite{18}. Goppa codes are of interest for two reasons: they have a good proved
minimum distance and  an efficient hard-decoding algorithm:
the Berlekamp-Massey algorithm for Goppa codes. Indeed, since the introduction of code-based
cryptography by McEliece in 1978 \cite{9}, Goppa codes still remain among the few families of algebraic codes which resist to any structural attack. This is one of the reasons
why every improvement of our knowledge of these codes is of particular interest.

The main practical limitation of the McEliece
public-key encryption scheme is probably the size of its key.
A famous trend to overcome this issue is to focus on subclasses of
Goppa codes with a non-trivial automorphism group. Such codes display then symmetries allowing compact parity-check or generator matrices. For instance, a key-reduction is obtained by taking quasi-cyclic (QC) or quasi-dyadic (QD) Goppa codes. Quasi-cyclic (QC)
codes represent a good example of the use of symmetries
in cryptography to build public-key encryption schemes with
short keys \cite{T,G}. It was then followed by a series of papers
proposing Alternant and Goppa codes with different automorphism groups like quasi-dyadic (QD) Goppa or Srivastava
codes \cite{M}, \cite{P} and quasi-monoidic (QM) codes \cite{B}. Besides these cryptographic
motivations, the search for Goppa codes, parity-check subcodes of Goppa codes, extended Goppa codes and  more general
Alternant codes, with non-trivial automorphisms is in itself an
important issue in coding theory. Several papers focused on the
problem of constructing quasi-cyclic Goppa codes \cite{BO, R}. Alternant
codes are subfield subcodes of Generalized Reed-Solomon codes and classical Goppa
codes are a special case of Alternant codes \cite{16}. Berger introduced cyclic Alternant codes induced by an automorphism of a GRS code in \cite{7}.  In \cite{6}, Berger proved that the
parity-check subcodes of Goppa codes and the extended Goppa codes are both Alternant
codes and gave new families of Goppa codes with a cyclic extension, and some families of
non-cyclic Goppa codes with a cyclic parity-check subcode. Moreover, he also identified Alternant and Goppa codes invariant under a
given permutation \cite{2}.  In \cite{F}, the authors constructed folding Alternant and Goppa Codes with
non-trivial automorphism groups.

In this paper,  we will present a sufficient and necessary condition for an irreducible monic polynomial $g(x)$ of degree $r$ over $\mathbb{F}_{q}$ satisfying $\gamma g(x)=(x+d)^rg({A}(x))$, where $A=\left(\begin{array}{cc} a&b\\1&d\end{array}\right)\in PGL_2(\Bbb F_{q})$, $q=2^n$, $\mathrm{ord}(A)$ is a prime, $g(a)\ne 0$  and  $0\ne \gamma\in \Bbb F_q$. And we shall determine  concrete form of  irreducible polynomials $g(x)$ of degree $2s$ or $3s$ as above, where $s$ is a positive integer. Moreover, we will give the constructions of binary irreducible quasi-cyclic parity-check subcodes of Goppa codes and extended Goppa codes.

The paper is organized as follows. In Section 2, we remind some definitions of
Alternant codes, Goppa codes,  parity-check subcodes  of Goppa codes and extended  Goppa codes.  In Section 3,  we present a sufficient and necessary condition for an  irreducible monic polynomial $g(x)$ of degree $r$ over $\mathbb{F}_{q}$. Moreover, we give a complete characterization of irreducible polynomials $g(x)$ of degree $2s$ or $3s$ as above, where $s$ is a positive integer. In Section 4,
we construct some binary irreducible quasi-cyclic parity-check subcodes of Goppa codes and extended Goppa codes.  We conclude the paper in Section 5.
\section{Preliminaries}
In this paper, we always assume that  $q=2^n$, $\mathbb{F}_{q}$ is the finite field of order $q$, and $\overline{\mathbb{F}}_{q}=\Bbb F_{q}\cup\{\infty\}$ is  a set of coordinates for the projective line. We will introduce some basic knowledge in the following.
\subsection{Alternant codes, Goppa codes, parity-check subcodes
of Goppa codes and  extended Goppa codes.}
In the subsection, we describe concepts of Alternant codes, Goppa codes, parity-check subcodes
of Goppa codes and extended Goppa codes. In detail,  see \cite{6}.

\begin{defn}
Let $L=(\alpha_0,\ldots,\alpha_{m-1})$ be an $m$-tuple of distinct elements of $\mathbb{F}_{q}$,  $v=(v_0,\ldots,v_{m-1})$  an $m$-tuple of non-zero elements of $\mathbb{F}_{q}$, and  $r$  an integer less than $m$. The Alternant code $\mathcal{A}_{r}(v,L)$ is defined as follows:
 $$\mathcal{A}_{r}(v,L)=\{x=(x_0,\ldots,x_{m-1})\in \mathbb{F}_{2}^m : H_r(v,L)x^T=0\},$$
 where the  parity-check matrix is
$$H_r(v,L)=\left(
  \begin{array}{cccc}
    v_0 & v_1 & \ldots & v_{m-1} \\
    v_0\alpha_0 & v_1\alpha_1 & \ldots & v_{m-1}\alpha_{m-1} \\
    \vdots& \vdots &  & \vdots \\
    v_0\alpha_0^{r-1} & v_1\alpha_1^{r-1} & \ldots & v_{m-1}\alpha_{m-1}^{r-1} \\
  \end{array}
\right).$$
\end{defn}


\begin{defn}\label{suppp}
Let $L=(\alpha_0,\ldots,\alpha_{m-1})$ be an $m$-tuple of distinct elements of $\mathbb{F}_{q}$ and $g(x)\in \mathbb{F}_{q}[x]$  a polynomial of degree $r(\leq n)$ such that $g(\alpha_i)\neq 0$ for $i=0,1,\ldots,m-1$. The Goppa code $\Gamma(g,L)$  with the  Goppa polynomial $g(x)$ and the support $L$ is defined as follows:
$$\Gamma(g,L)=\{x=(x_0,x_2,\ldots, x_{m-1})\in \Bbb F_2^m: \sum_{i=0}^{m-1}\frac {x_i}{x-\alpha_i}\equiv 0\pmod {g(x)}\}.$$
If $g(x)$ is irreducible over $\mathbb{F}_{q}$,  $\Gamma(g,L)$ is called irreducible.
\end{defn}

\begin{defn}
The parity-check subcode  $\widetilde{\Gamma}(g,L)$ of  $\Gamma (L, g)$
is defined as follows:
$$\widetilde{\Gamma}(g,L)=\{x=(x_0,\ldots,x_{m-1})\in \Gamma(g,L): \sum\limits_{i=0}^{m-1}x_i=0\}.$$

The extended Goppa code $\overline{\Gamma}(L,g)$ is defined as follows:
$$\overline{\Gamma}(g,L)=\{x=(x_0,\ldots,x_{m-1},x_m): (x_0,\ldots,x_{m-1})\in \Gamma(g,L), \sum\limits_{i=0}^{m}x_i=0\}.$$
\end{defn}
\begin{rem} Let $\Gamma(g,L)$ be the Goppa code in Definition \ref{suppp}.  Then

(1)  $\Gamma(g, L)=\mathcal{A}_r(v_{g,L},L)$, where  $v_{g,L}=(g(\alpha_0)^{-1},g(\alpha_1)^{-1},\ldots,g(\alpha_{m-1})^{-1})$;

(2) $\widetilde{\Gamma}(g,L)=\mathcal A_{r+1}(v_{g,L}, L)$, where  $v_{g,L}=(g(\alpha_0)^{-1},g(\alpha_1)^{-1},\ldots,g(\alpha_{m-1})^{-1})$.

\end{rem}

\begin{defn}
Let $L=(\alpha_0,\ldots,\alpha_{m-1})$ be an $m$-tuple of distinct elements of $\mathbb{F}_{q}$, $\overline{L}=L \bigcup \{\infty\}=(\alpha_0,\ldots,\alpha_{m-1},\infty)$, $v=(v_0,\ldots,v_m)$ an $(m+1)$-tuple of non-zero elements of $\mathbb{F}_{q}$, and $r$ an integer less than $m+1$. The Alternant code $\mathcal{A}_r(v,\overline{L})$ is defined as follows:
 $$\mathcal{A}_r(v,\overline{L})=\{x=(x_0,\ldots,x_{m})\in \mathbb{F}_2^{m+1}: H_r(v,\overline{L})x^T=0\},$$
where the parity-check matrix is
$$H_r(v,\overline{L})=\left(
  \begin{array}{cccc}
    v_0 & \ldots & v_{m-1} & 0 \\
    \vdots & & \vdots & \vdots \\
    v_0\alpha_0^{r-2} & \ldots & v_{m-1}\alpha_{m-1}^{r-2}& 0 \\
    v_0\alpha_0^{r-1} & \ldots & v_{m-1}\alpha_{m-1}^{r-1}& v_m \\
  \end{array}
\right).$$
\end{defn}


\begin{rem}\label{extn} Let $\Gamma(g,L)$ be the Goppa code in Definition 2.2,
 $g(x)=\sum\limits_{i=0}^{r}g_ix^i$  a polynomial of degree $r$, and  $\overline L=L\cup \{\infty\}$.
Then the extended Goppa code $\overline{\Gamma}(g,L)$ of $\Gamma(g, L)$ is just the Alternant code $\mathcal{A}_{r+1}(v_{g,\overline{L}},\overline{L})$, where $v_{g,\overline{L}}=(g(\alpha_0)^{-1},\ldots,g(\alpha_{m-1})^{-1},g(\infty)^{-1})$, $g(\infty)=g_r$.
\end{rem}

\subsection{Action of  groups}\label{agl}
We will recall the actions of the projective linear group  and the projective semi-linear group on $\mathbb{F}_{q}$ and $\overline{\mathbb{F}}_{q}=\Bbb F_{q}\cup \{\infty\}$, respectively.

There are some matrix  groups as follows:

(1) The affine group $$AGL_2(\mathbb{F}_{q})=\{A=\left(\begin{array}{cc}a &b \\ 0 & 1\end{array}\right):   a \in \mathbb{F}_{q}^*, b \in \mathbb{F}_{q}\}.$$

(2) The general linear group
$$GL_2(\mathbb{F}_{q})=\left\{A=\left(
                               \begin{array}{cc}
                                 a & b \\
                                 c & d \\
                               \end{array}
                             \right)
 : a,b,c,d \in \mathbb{F}_{q}, ad-bc \neq 0\right\}.$$

(3) The  projective linear group
$$PGL_2(\mathbb{F}_{q})=GL_2(\Bbb F_{q})/\{aE_2: a\in \Bbb F_{q}^*\},$$
where $E_2$ is the $2\times 2$ identity  matrix.

(4) The projective semi-linear group
$$P\Gamma L_2(\Bbb F_{q})=PGL_2(\Bbb F_{q})\times G,$$
where $G=Gal(\Bbb F_{q}/\Bbb F_p)=\langle \sigma\rangle$ is the Galois group.

Let $\overline {\Bbb F}_{q}=\Bbb F_{q}\cup \{\infty\}$ be the  projective line set and  $A=\left(\begin{array}{cc} a&b\\c&d\end{array}\right)\in GL_2(\Bbb F_{q})$. Then the projective linear group $PGL_2(\Bbb F_{q})$ acts on $\overline {\Bbb F}_{q}$ as follows:
\begin{eqnarray*}PGL_2(\Bbb F_{q})\times \overline{\Bbb F}_{q}&\rightarrow& \overline {\Bbb F}_{q}\\
(\widetilde{A}, \zeta)&\mapsto& \widetilde A(\zeta)=A(\zeta)=\frac{a\zeta+b}{c\zeta+d},\end{eqnarray*}
where
$\frac{1}{0}=\infty$ and  $\frac{1}{\infty}=0$.

 The projective semi-linear group $P\Gamma L_2(\Bbb F_{q})$ acts on $\overline {\Bbb F}_{q}$ as follows:
\begin{eqnarray*}P\Gamma L_2(\Bbb F_{q})\times \overline{\Bbb F}_{q}&\rightarrow& \overline {\Bbb F}_{q}\\
((\widetilde{A},\sigma^i), \zeta)&\mapsto& (\widetilde A, \sigma^i)(\zeta)=A(\sigma^i(\zeta))=\frac{a\zeta^{p^i}+b}{c\zeta^{p^i}+d}.\end{eqnarray*}


In the following, if $A=\left(\begin{array}{cc} a&b\\c&d\end{array}\right)\in PGL_2(\Bbb F_{q})(c\neq0)$, then we use the convention $c = 1$.

 \section{ Characterization of  polynomial $g(x)$}\label{agl}
In this section, we characterize the form of $g(x)$ satisfying $\gamma g(x)=(x+d)^rg({A}(x))$, where $A=\left(\begin{array}{cc} a&b\\1&d\end{array}\right)\in PGL_2(\Bbb F_{q})$, $g(a)\ne 0$, and $0\ne \gamma\in \Bbb F_q$. In some conditions, we present a sufficient and necessary condition for an   irreducible monic  polynomial $g(x)$. Moreover, we give a complete characterization of irreducible polynomials $g(x)$.
\begin{lem}\label{c11}
Let $g(x)$ be a monic polynomial of degree $r$ over $\mathbb{F}_{q}$, $A=\left(\begin{array}{cc} a&b\\1&d\end{array}\right)\in PGL_2(\Bbb F_{q})$, and $g(a)\ne 0$. Then $\gamma g(x)=(x+d)^rg({A}(x))$,  $0\ne \gamma\in \Bbb F_q$, if and only if
$$g(x)=\prod^{s}\limits_{j=1}\prod^{l_j-1}\limits_{i=0}(x-A^i(\beta_j)),$$
where $r=\sum^{s}\limits_{j=1}l_j$ and each $l_j$ is the least positive integer such that $A^{l_j}(\beta_j)=\beta_j$.
\end{lem}

\begin{proof} Suppose that ${\overline g}(x)=(x+d)^rg({A}(x))$. Then $\deg(\overline {g}(x))=r$ by $\deg(g(x))=r$ and $g(a)\ne 0$.
Hence
   $\overline g(x)=\gamma g(x)$, $0\ne \gamma\in \Bbb F_q$, is equivalent to  that    $g(A(\beta))=0$  if $g(\beta)=0$, which is equivalent to
 that  $g(x)=\prod^{s}\limits_{j=1}\prod^{l_j-1}\limits_{i=0}(x-A^i(\beta_j))$, where each $l_j$ is the least positive integer such that $A^{l_j}(\beta_i)=\beta_j$.
\end{proof}
\begin{lem}\label{lem} Let $A(\beta)=\frac{a\beta +b}{\beta +d}=\beta$, where
$A=\left(\begin{array}{cc} a&b\\1&d\end{array}\right)\in PGL_2(\Bbb F_{q})$  and $\beta$ is an element in an extension over $\Bbb F_q$. Then $\beta\in \mathbb{F}_{q}$ if $a=d$; $\beta\in \mathbb{F}_{q^2}$
if $a\neq d$.
\end{lem}
\begin{proof}
Let $A(\beta)=\frac{a\beta +b}{\beta +d}=\beta$, then
 \begin{equation*}
   b=
   \beta^2+(a+d)\beta.
 \end{equation*}

 If $a=d$, then $b=\beta^2$ and $\beta\in \Bbb F_q$.

 If $a\ne d$, then $\beta^2+(a+d)\beta+b=0$ and $\beta\in \Bbb F_{q^2}$.

 The proof of this lemma is done.
\end{proof}

\begin{thm}\label{th1} Let $g(x)$ be a monic polynomial of degree $r$ over $\Bbb F_q$ such that $\gamma g(x)=(x+d)^rg(A(x))$, where $0\ne \gamma\in \Bbb F_q$, $A=\left(\begin{array}{cc} a&b\\1&d\end{array}\right)\in PGL_2(\Bbb F_{q})$, $\ord(A)=l$ is a prime, and $g(a)\ne 0$. Then $g(x)$ is irreducible over $\Bbb F_q$ if and only if $$g(x)=\prod^{s-1}\limits_{j=0}\prod^{l-1}\limits_{i=0}(x-A^i(\beta^{q^j})),$$ where $r=sl$, $s$ is the least positive integer such that $A^u(\beta)=\beta^{q^s}$, $1\le u\le l-1$.
\end{thm}
\begin{proof} For $\beta$ in an extension over $\Bbb F_q$,  let $l'$ be the least positive integer such that $A^{l'}(\beta)=\beta$. Then $l'=1$ or $l$ by $\ord(A)=l$ prime.

Suppose that $g(x)$ is irreducible over $\Bbb F_q$. Then by Lemmas 3.1 and 3.2, $g(x)=\prod^{s}\limits_{j=1}\prod^{l-1}\limits_{i=0}(x-A^i(\beta_j))$
 and $r=sl$. Let $\beta=\beta_1$ and $A(\beta)=\beta^{q^a}$ by Galois theory. Then $\beta=A^l(\beta)=\beta^{q^{al}}$ and $sl|al$, so $a=sk$, $k\in \Bbb Z$, and $\gcd(k,l)=1$. Hence there are $u,v \in \Bbb Z$ such that $uk+vl=1$ and $A^u(\beta)=\beta^{q^{sku}}=\beta^{q^s}$, $1\le u\le l-1$. Set $\beta_j=\beta^{q^{j-1}}$, $1\le j\le s$,  and  by Galois theory there is the factorization of $g(x)$.

Conversely, by $s$ is the least positive integer such that $A^u(\beta)=\beta^{q^s}$,  $\{A^i(\beta^{q^j})| 0\le i\le l-1,  0\le j\le s-1\}=\{\beta^{q^{is+j}}|0\le i\le l-1, 0\le j\le s-1\}$. By Galois theory, $g(x)$ is irreducible over $\Bbb F_q$.
\end{proof}
For $l$  a prime,
define
\begin{equation}N_l(\Bbb F_q)=\{{A}=\left(\begin{array}{cc} a&b\\1&d\end{array}\right)\in PGL_2(\mathbb{F}_{q}) \mid \ord(A)=l\}.\end{equation}

In the following, we find the sets $N_2(\Bbb F_q)$ and $N_3(\Bbb F_q)$.

\begin{lem}\label{le3} (1) $$N_2(\Bbb F_q)=\{ A=\left(\begin{array}{cc} a&b\\1&a\end{array}\right)\mid a, b\in \Bbb F_q, a^2\ne b\}$$
and  $|N_2(\Bbb F_q)|=q(q-1)$.

(2) $$N_3(\Bbb F_q)=\{ A=\left(\begin{array}{cc} a&a^2+ad+d^2\\1&d\end{array}\right)\mid  a\ne d,  a,d\in\Bbb F_q\}$$
and $|N_3(\Bbb F_q)|=q(q-1)$.

If $a=0$ in $N_3(\Bbb F_q)$, then
$$N_{3,a}(\Bbb F_q)=\{ A=\left(\begin{array}{cc} 0&d^2\\1&d\end{array}\right)\mid 0\ne d\in\Bbb F_q\}$$
and $|N_{3,a}(\Bbb F_q)|=q-1$.

 If $d=0$ in $N_3(\Bbb F_q)$, then
$$N_{3,d}(\Bbb F_q)=\{ A=\left(\begin{array}{cc} a&a^2\\1&0\end{array}\right)\mid 0\ne a\in\Bbb F_q\}$$
and $|N_{3,d}(\Bbb F_q)|=q-1$.

If $b=0$ in $N_3(\Bbb F_q)$ and $3|(q-1)$, then
$$N_{3,b}(\Bbb F_q)=\{ A=\left(\begin{array}{cc} a&0\\1&a\omega^i\end{array}\right)\mid 0\ne a\in\Bbb F_q, i=1,2\}$$
and $|N_{3,b}(\Bbb F_q)|=2(q-1)$, where $\omega\in \Bbb F_q$ and $\ord(\omega)=3$.

\end{lem}
\begin{proof} For   $A=\left(\begin{array}{cc} a&b\\1&d\end{array}\right)\in PGL_2(\Bbb F_q)$,
 ${A}^2=\left(\begin{array}{cc} a^2+b&b(a+d)\\a+d&b+d^2\end{array}\right)$. Then  $ \ord({A})=2$ if and only if $a=d$ and $a^2\neq b$. Hence
$$N_2(\Bbb F_q)=\{ A=\left(\begin{array}{cc} a&b\\1&a\end{array}\right)\mid a, b\in \Bbb F_q, a^2\ne b\}$$
and  $|N_2(\Bbb F_q)|=q(q-1)$.

 On the other hand, ${A}^3=\left(\begin{array}{cc} a^3+bd&b(a^2+ad+d^2+b)\\a^2+ad+d^2+b&d^3+ab\end{array}\right)$. Then  $\ord(A)=3$ if and only if
\begin{equation*}\label{eqn_system_3}
\left\{
\begin{array}{l}
a^2+ad+d^2+b=0\\
a^3+bd=d^3+ab\neq0\\
ad\neq b\\
a\neq d
\end{array} \right.\ \
\end{equation*}
if and only if  $$ \left\{\begin{array}{l}a^2+ad+d^2+b=0\\ a\neq d.\end{array}\right.$$
Hence
$$N_3(\Bbb F_q)=\{ A=\left(\begin{array}{cc} a&a^2+ad+d^2\\1&d\end{array}\right)\mid  a\ne d,  a,d\in\Bbb F_q\}$$
and $|N_3(\Bbb F_q)|=q(q-1)$.

If $a=0$, then
\begin{equation}N_{3,a}(\Bbb F_q)=\{ A=\left(\begin{array}{cc} 0&d^2\\1&d\end{array}\right)\mid  0\ne d\in\Bbb F_q\}\end{equation}
and $|N_{3,a}(\Bbb F_q)|=q-1$.

If $d=0$, then $$N_{3,d}(\Bbb F_q)=\{ A=\left(\begin{array}{cc} a&a^2\\1&0\end{array}\right)\mid  0\ne a\in\Bbb F_q\}$$
and $|N_{3,d}(\Bbb F_q)|=q-1$.

If $b=0$ and $3|(q-1)$, then by $a^2+ad+d^2=0$, $d/a=\omega^i$ and $d=a\omega^i$, $i=1,2$, where $\omega\in \Bbb F_q$ and  $\ord(\omega)=3$.
Hence $$N_{3,b}(\Bbb F_q)=\{ A=\left(\begin{array}{cc} a&0\\1&a\omega^i\end{array}\right)\mid 0\ne a\in\Bbb F_q, i=1,2\}$$
and $|N_{3,b}(\Bbb F_q)|=2(q-1)$, where $\omega\in \Bbb F_q$ and $\ord(\omega)=3$.

The proof of this lemma is done.
\end{proof}

Let $A\in PGL_2(\Bbb F_q)$ and $\ord(A)=l=2$ or $3$. In the following, we   give a complete characterization of irreducible polynomial $g(x)=\prod^{s-1}\limits_{j=0}\prod^{l-1}\limits_{i=0}(x-A^i(\alpha^{q^j}))$ over $\mathbb{F}_{q}$, where $\alpha$ is an element in an extension $\Omega$ over $\Bbb F_q$.

 In the following, we always assume that $\Omega$ is an extension over $\Bbb F_q$ and $\Tr^{sn}_s(\cdot)$ is the trace function from $\mathbb{F}_{2^{sn}}$ to $\mathbb{F}_{2^s}$, i.e., for $x\in\Bbb F_{2^{sn}}$, $\Tr^{sn}_{s}(x)=\sum^{n-1}\limits_{i=0}x^{2^{si}}$.

\begin{lem}{\cite{H}}
If $t\in\mathbb{F}_{q}$, the quadratic equation $x^2+x=t$ has solutions in $\mathbb{F}_{q}$ if and only if $\Tr^{n}_{1}(t)=0$.
\end{lem}
\begin{thm}{\label{th6}}
Let $A=\left(\begin{array}{cc} a&b\\1&a\end{array}\right)\in PGL_2(\mathbb{F}_{q})$ be a matrix of order $2$.
Then
 $$g_k(x)=x^2+kx+ak+b=(x-\alpha)(x-A(\alpha)), \alpha\in \Omega$$ is irreducible over $\Bbb F_q$ if and only if
  $k\in T$, where $T=\{\frac{a+\sqrt b}c:\Tr^n_1(c)=1, c\in \Bbb F_q\}$. In fact,  $q$ is the least positive integer such that $A(\alpha)=\alpha^q$.

Moreover, there is an irreducible factorization of $h(x)=x^{q+1}+ax^{q}+ax+b$ over $\Bbb F_q$:
 $$h(x)=(x-\sqrt{b})\prod\limits_{k\in{T}}g_k(x).$$
\end{thm}
\begin{proof}
Suppose that  $g_k(x)=x^2+kx+ak+b$ and $k\ne0$. Then by $x=yk$,  $$g_k(yk)=k^2y^2+k^2y+ak+b$$
 and
$$\frac{g_k(yk)}{k^2}=y^2+y+\frac{ak+b}{k^2}=(y+\frac {\sqrt b}k)^2+(y+\frac{\sqrt b}k)+\frac {a+\sqrt b}k.$$

Hence by Lemma 3.5,  $g_k(x)$ is irreducible over $\Bbb F_q$ if and only if $\Tr^n_1(\frac {a+\sqrt b}k)=1$ if and only if $k\in T=\{\frac {a+\sqrt b}{c}: \Tr^n_1(c)=1, c\in \Bbb F_q\}$.  If $k=0$, then $g_k(x)$ is reducible.

In fact, let $\alpha_1=\alpha$ and $\alpha_2$ be two roots of the irreducible polynomial  $g_k(x)$ over $\Bbb F_q$, then
$k=\frac{\alpha^2+b}{\alpha +a}$ and $\alpha_2=\frac{a\alpha+b}{\alpha+a}=A(\alpha)$, so $g_k(x)=(x-\alpha)(x-A(\alpha))$ and $A(\alpha)=\alpha^q$.

If $k\in T$ and $\alpha$ is a root of the irreducible polynomial $g_k(x)$, then $A(\alpha)=\alpha^q$ and $\alpha$ is a root of $h(x)=x^{q+1}+ax^q+ax+b$, so $g_k(x)|h(x)$. It is clear that $\sqrt b$ is also a root of $h(x)$.
If $k_1\ne k_2\in T$, then $\gcd(g_{k_1}(x),g_{k_2}(x))=1$ and $|T|=\frac{q}2$.
Hence
$$h(x)=x^{q+1}+ax^{q}+ax+b=(x-\sqrt b)\prod\limits_{k\in{T}}g_k(x).$$
 The proof of this theorem is done.
\end{proof}

Now we consider Theorem 3.3 in the case $r=2s$.

\begin{thm}{\label{th7}}
Let $A=\left(\begin{array}{cc} a&b\\1&a\end{array}\right)\in PGL_2(\mathbb{F}_{q})$ be a matrix of order $2$.

(1)  Suppose that $s=t^e$, where $t$ is an odd prime and $e$ is a positive integer.
Then for $0\leq i\leq e$,
$$g_{i,k}(x)=\prod^{t^i-1}\limits_{j=0}(x^2+k^{q^j}x+ak^{q^j}+b)=\prod^{t^i-1}\limits_{j=0}(x-\alpha^{q^j})(x-A(\alpha^{q^j})), \alpha\in\Omega$$
is irreducible over $\Bbb F_q$ if and only if $k\in {T^{(i)}}\cap(\Bbb F_{q^{t^i}}\backslash\Bbb F_{q^{t^{i-1}}})$, where $\Bbb F_{q^{t^{-1}}}=\emptyset$,  ${T^{(i)}}=\{\frac{a+\sqrt b}{c}: \Tr^{t^in}_1(c)=1, c\in\Bbb F_{q^{t^i}}\}$.
In fact, $q^{t^i}$ is the least positive integer such that  $A(\alpha)=\alpha^{q^{t^i}}$.

Moreover, there is an irreducible factorization of $h(x)=x^{q^s+1}+ax^{q^s}+ax+b$ over $\Bbb F_q$:
 \begin{equation}h(x)=(x-\sqrt{b})\prod_{i=0}^e\prod\limits_{k\in({T^{(i)}}\cap(\Bbb F_{q^{t^i}}\backslash\Bbb F_{q^{t^{i-1}}}))/\thicksim}g_{i,k}(x),\end{equation}
where  $({T^{(i)}}\cap(\Bbb F_{q^{t^i}}\backslash\Bbb F_{q^{t^{i-1}}}))/\thicksim$ is a quotient set, and  $\sim$ is an equivalent relation of a set ${T^{(i)}}\cap(\Bbb F_{q^{t^i}}\backslash\Bbb F_{q^{t^{i-1}}})$: for $k', k\in {T^{(i)}}\cap(\Bbb F_{q^{t^i}}\backslash\Bbb F_{q^{t^{i-1}}})$, $k'\sim k$ if and only if $k'=k^{q^j}, 0\le j\le t^i-1$.

(2)  Suppose that  $s=t_1t_2$, where $t_1, t_2$ are two distinct odd primes. Then
 $$g_{k}(x)=x^2+kx+ak+b=(x-\alpha)(x-A(\alpha)),  \alpha\in\Omega$$  is irreducible over $\Bbb F_{q}$ if and only if  $k\in{T^{(0)}}$, where ${T^{(0)}}=\{\frac{a+\sqrt b}c:\Tr^{n}_1(c)=1, c\in \Bbb F_{q}\}$. In fact, $q$ is the least positive integer such that  $A(\alpha)=\alpha^{q}$.

For $t_i(i=1,2)$,
 $$g_{t_{i}, k}(x)=\prod^{t_i-1}\limits_{j=0}(x^2+k^{q^j}x+ak^{q^j}+b)=\prod_{j=0}^{t_i-1}(x-\alpha^{q^j})(x-A(\alpha^{q^j})), \alpha\in\Omega$$ is an irreducible polynomial over $\Bbb F_q$
 if and only if  $k\in T^{(t_i)}\cap(\Bbb F_{q^{t_i}}\backslash\Bbb F_{q})$, where $T^{(t_i)}=\{\frac{a+\sqrt b}c: \Tr^{t_in}_1(c)=1, c\in \Bbb F_{q^{t_i}}\}$. In fact,  $q^{t_i}$ is the least positive integer such that  $A(\alpha)=\alpha^{q^{t_i}}$.

For $s=t_1t_2$,
  $$g_{s, k}(x)=\prod^{s-1}\limits_{j=0}(x^2+k^{q^j}x+ak^{q^j}+b)=\prod^{s-1}\limits_{j=0}(x-\alpha^{q^j})(x-A(\alpha^{q^j})), \alpha\in\Omega$$  is irreducible over $\Bbb F_q$
 if and only if
  $k\in T^{(s)}\cap(\Bbb F_{q^s}\backslash(\Bbb F_{q^{t_1}}\cup\Bbb F_{q^{t_2}}))$, where $T^{(s)}=\{\frac{a+\sqrt b}c:\Tr^{sn}_1(c)=1, c\in \Bbb F_{q^s}\}$. In fact, $q^s$ is the least positive integer such that  $A(\alpha)=\alpha^{q^{s}}$.

Moreover, there is an irreducible factorization of $h(x)=x^{q^s+1}+ax^{q^s}+ax+b$ over $\Bbb F_q$:
  \begin{eqnarray}h(x)&=&(x-\sqrt{b})\prod\limits_{k\in{T^{(0)}}}g_{k}(x)\prod\limits_{k\in({T^{(t_1)}}\cap(\Bbb F_{q^{t_1}}\backslash\Bbb F_{q}))/\sim}g_{t_1,k}(x)\\
 &&\prod\limits_{k\in({T^{(t_2)}}\cap(\Bbb F_{q^{t_2}}\backslash\Bbb F_{q}))/\sim}g_{t_2,k}(x)
 \prod\limits_{k\in{(T^{(s)}}\cap(\Bbb F_{q^s}\backslash(\Bbb F_{q^{t_1}}\cup\Bbb F_{q^{t_2}})))/\sim}g_{s,k}(x),\nonumber
 \end{eqnarray}
where all  $\sim$ are equivalent relations as above.
\end{thm}
\begin{proof} (1) Suppose that $s=t^e$, where $t$ is an odd prime and $e$ is a positive integer. Let $\alpha$ be a root of $h(x)=x^{q^s+1}+ax^{q^s}+ax+b$ in an extension over $\Bbb F_q$. Then $A(\alpha)=\alpha^{q^s}$ and $\alpha=\alpha^{q^{2s}}$.
Let $l$ be the least non-negative integer such that $A(\alpha)=\alpha^{q^l}$ and $\alpha=\alpha^{q^{2l}}$. If $l=0$, then $\alpha=\sqrt b$.

If $l\ne 0$, then $l|s$ and $l=t^i$, $0\le i\le e$.
Hence  by   Theorem \ref{th6},
$x^2+kx+ak+b=(x-\alpha)(x-A(\alpha))$  is irreducible over $\Bbb F_{q^{t^i}}$ if and only if  $k\in{T^{(i)}}$, where ${T^{(i)}}=\{\frac{a+\sqrt b}c:\Tr^{t^in}_1(c)=1, c\in \Bbb F_{q^{t^i}}\}$. Since $q^{t^i}$ is the least positive integer such that $A(\alpha)=\alpha^{q^{t^i}}$, $k\in \Bbb F_{q^{t^i}}\setminus \Bbb F_{q^{t^{i-1}}}$.
Hence
 $[\Bbb F_q(\alpha):\Bbb F_q]=2t^i$ if and only if $k\in{T^{(i)}}\cap(\Bbb F_{q^{t^i}}\backslash\Bbb F_{q^{t^{i-1}}})$ if and only if $$g_{i,k}(x)=\prod^{t^i-1}\limits_{j=0}(x^2+k^{q^j}x+ak^{q^j}+b)=\prod_{j=0}^{t^i-1}(x-\alpha^{q^j})(x-A(\alpha^{q^j}))$$ is an irreducible polynomial over $\Bbb F_q$.
In fact, for $k',k\in{T^{(i)}}\cap(\Bbb F_{q^{t^i}}\backslash\Bbb F_{q^{t^{i-1}}})$, $k'\sim k$ if and only if   $k'=k^{q^j}$, $0\le j\le t^i-1$,  if and only if  $g_{i, k}(x)=g_{i,k'}(x)$, where $\sim$ is an equivalent relation. So there is a quotient set $({T^{(i)}}\cap(\Bbb F_{q^{t^i}}\backslash\Bbb F_{q^{t^{i-1}}}))/\sim$.

Note that $\sqrt b$ is  a root of $h(x)=x^{q^s+1}+ax^{q^s}+ax+b$.
  For $k\in({T^{(i)}}\cap(\Bbb F_{q^{t^i}}\backslash\Bbb F_{q^{t^{i-1}}}))/\sim$, $g_{i,k}(x)=\prod^{t^i-1}\limits_{j=0}(x-\alpha^{q^j})(x-A(\alpha^{q^j}))$ is irreducible over $\Bbb F_q$ and
  $q^{t^i}$ is the least  positive such that  $A(\alpha)=\alpha^{q^{t^i}}$.
 Then $g_{i,k}(x)|h(x)$, $0\le i\le e$.

  By $t$ is  an odd prime and $1\le i\le e$,  $\Tr^{t^in}_{t^{i-1}n}(\Bbb F_{q^{t^i}})=\Bbb F_{q^{t^{i-1}}}$ and $|{T^{(i)}}\cap(\Bbb F_{q^{t^i}}\backslash\Bbb F_{q^{t^{i-1}}})|=\frac {q^{t^i}-q^{t^{i-1}}}2$. Then  $\sum^e\limits_{i=0}|{T^{(i)}}\cap(\Bbb F_{q^{t^i}}\backslash\Bbb F_{q^{t^{i-1}}})|=\frac q 2+\sum^e\limits_{i=1}\frac {q^{t^i}-q^{t^{i-1}}}2=\frac {q^s}2$. Hence there is an irreducible factorization of $h(x)$ over $\Bbb F_q$ in (3.3).

(2) Suppose that $s=t_1t_2$, where $t_1$ and $t_2$ are  distinct odd primes. Then $\Bbb F_{q^s}$ has four distinct subfields over $\Bbb F_q$: $\Bbb F_q$, $\Bbb F_{q^{t_1}}$, $\Bbb F_{q^{t_2}}$, $\Bbb F_{q^s}$.  Let $\alpha$ be a root of $h(x)=x^{q^s+1}+ax^{q^s}+ax+b$ in an extension $\Omega$ over $\Bbb F_q$. Then $A(\alpha)=\alpha^{q^s}$ and $\alpha=\alpha^{q^{2s}}$. Let $l$ be the least non-negative integer such that $A(\alpha)=\alpha^{q^l}$ and $\alpha=\alpha^{q^{2l}}$. If $l=0$, then $\alpha=\sqrt b$.
If $l\ne0$, then $l|s$ and $l\in\{1, t_1, t_2, s\}$.

If $l=1$, then  by   Theorem \ref{th6},
$g_{k}(x)=x^2+kx+ak+b=(x-\alpha)(x-A(\alpha))$  is irreducible over $\Bbb F_{q}$ if and only if  $k\in{T^{(0)}}$, where $T^{(0)}=\{\frac{a+\sqrt b}c:\Tr^{n}_1(c)=1, c\in \Bbb F_{q}\}$. And $q$ is the least positive integer such that $A(\alpha)=\alpha^q$.

If $l=t_i$, $i=1, 2$, then  by (1) of Theorem 3.7, $$g_{t_{i}, k}(x)=\prod^{t_i-1}\limits_{j=0}(x^2+k^{q^j}x+ak^{q^j}+b)=\prod_{j=0}^{t_i-1}(x-\alpha^{q^j})(x-A(\alpha^{q^j}))$$ is an irreducible polynomial over $\Bbb F_q$
 if and only if  $k\in{T^{(t_i)}}\cap(\Bbb F_{q^{t_i}}\backslash\Bbb F_{q})$, where ${T^{(t_i)}}=\{\frac{a+\sqrt b}c: \Tr^{t_in}_1(c)=1, c\in \Bbb F_{q^{t_i}}\}$. And  $t_i$ is the least positive integer such that  $A(\alpha)=\alpha^{q^{t_i}}$.
In fact, for $k',k\in{T^{(t_i)}}\cap(\Bbb F_{q^{t_i}}\backslash\Bbb F_{q})$, $k'\sim k$ if and only if   $k'=k^{q^j}$, $0\le j\le t_i-1$,  if and only if  $g_{t_i, k}(x)=g_{t_i,k'}(x)$, where $\sim$ is an equivalent relation. So there is a quotient set $({T^{(t_i)}}\cap(\Bbb F_{q^{t_i}}\backslash\Bbb F_{q}))/\sim$.

If $l=s$,  then  by   Theorem \ref{th6},
$x^2+kx+ak+b=(x-\alpha)(x-A(\alpha))$  is irreducible over $\Bbb F_{q^{s}}$ if and only if  $k\in T^{(s)}$, where ${T^{(s)}}=\{\frac{a+\sqrt b}c:\Tr^{sn}_1(c)=1, c\in \Bbb F_{q^{s}}\}$. Since $q^{s}$ is the least positive integer such that $A(\alpha)=\alpha^{q^{s}}$, $k\in \Bbb F_{q^{s}}\setminus (\Bbb F_{q^{t_1}}\cup \Bbb F_{q^{t_2}})$.
Hence
 $[\Bbb F_q(\alpha):\Bbb F_q]=2s$ if and only if $k\in{T^{(s)}}\cap(\Bbb F_{q^{s}}\backslash(\Bbb F_{q^{t_1}}\cup \Bbb F_{q^{t_2}})$ if and only if $$g_{s,k}(x)=\prod^{s-1}\limits_{j=0}(x^2+k^{q^j}x+ak^{q^j}+b)=\prod_{j=0}^{s-1}(x-\alpha^{q^j})(x-A(\alpha^{q^j}))$$ is an irreducible polynomial over $\Bbb F_q$.
Also there is a quotient set $({T^{(s)}}\cap(\Bbb F_{q^{s}}\backslash(\Bbb F_{q^{t_1}}\cup \Bbb F_{q^{t_2}}))/\sim$.

Note that $\sqrt b$ is  a root of $h(x)=x^{q^s+1}+ax^{q^s}+ax+b$. For $k\in T^{(0)}$, $g_{k}(x)|h(x)$.
  For $i=1,2$ and  $k\in({T^{(t_i)}}\cap(\Bbb F_{q^{t_i}}\backslash\Bbb F_{q}))/\sim$,  $g_{t_i, k}(x)|h(x)$.
 For $k\in (T^{(s)}\cap (\Bbb F_{q^{s}}\setminus(\Bbb F_{q^{t_1}}\cup\Bbb F_{q^{t_2}}))/\sim$,  $g_{s, k}|h(x)$.

  By $t_1$ and $t_2$ are  odd primes,  $|T^{(0)}|+\sum\limits_{i=1}^2|T^{(t_i)}\cap (\Bbb F_{q^{t_i}}\setminus \Bbb F_q))|+|(T^{(s)}\cap (\Bbb F_{q^{s}}\setminus(\Bbb F_{q^{t_1}}\cup\Bbb F_{q^{t_2}}))|=\frac q2+\frac {q^{t_1}-q}2+\frac{q^{t_2}-q}2+\frac{q^{s}-q^{t_1}-q^{t_2}+q}2=\frac{q^{s}}2$.
Hence there is an irreducible factorization of $h(x)$ over $\Bbb F_q$ in (3.4).

 The proof of this theorem is done.
\end{proof}

In \cite{Berlekamp}, Berlekamp et al. gave a sufficient and necessary conditions for the cubic equation $x^3+x=t$ $(t\in \mathbb{F}^*_{2^n})$ has a unique solution in $\mathbb{F}_{2^n}$. Furthermore, if there is not a unique solution, they do not know whether there are
no solutions or three solutions. Although  Berlekamp et al. provided a theoretical answer to the question of
whether $x^3+x=t$  has three roots or zero roots in \cite{Berlekamp1966}, it is not useful in practice. Hence, it is hard to know whether $g(x)=x^3+\sigma_1x^2+\sigma_2x+\sigma_3(\sigma_1\neq\sigma^2_2)$ is irreducible over $\mathbb{F}_{2^n}$. In the  following, we give  a method to  determine how many roots there are and give a complete characterization of irreducible polynomials of degree 3.

\begin{thm}{\label{th8}}
Let $3\mid(q-1)$, $\xi$ be a primitive element of $\mathbb{F}_{q}$, and   $\omega=\xi^{\frac{q-1}{3}}$  be an element of order $3$.  Without loss of generality, we assume that
 $\beta$ is  a root of the irreducible polynomial $x^3-\xi$ over $\Bbb F_q$ such that $\beta^q=\omega\beta$.
Let
$D_0=<\xi^3>$ be a cyclic multiplicative subgroup of $\mathbb{F}^*_{q}$ and   $D_j=\xi^j<\xi^3>$, $j=1, 2$,  two cosets.
Let $A=\left(\begin{array}{cc} a&a^2+ad+d^2\\1&d\end{array}\right)\in PGL_2(\mathbb{F}_{q})$ be a matrix of order $3$. Then
$$g_k(x)=x^3+kx^2+(a^2+k(a+d)+ad+d^2)x+a^3+kad+d^3=\prod^{2}\limits_{i=0}(x-A^i(\alpha)), \alpha\in\Omega,$$ is irreducible over $\mathbb{F}_{q}$ if and only if $k\in T_1\cup T_2$, where  $$T_j=\{\frac{a+d}{c+1}+a\omega+d\omega^2: c\in D_j\}, j=1,2.$$ In fact,  $q$ is the least positive integer such that $A^2(\alpha)=\alpha^q$ for $k\in T_1$;  $q$ is the least positive integer such that $A(\alpha)=\alpha^q$ for $k\in T_2$.

Moreover, there are two irreducible factorizations of $h(x)$ over $\Bbb F_q$:
$$h(x)=x^{q+1}+ax^q+dx+a^2+ad+d^2=(x+a\omega+d\omega^2)(x+a\omega^2+d\omega)\prod\limits_{k\in T_1}g_k(x),$$
 and
$$h(x)=x^{q+1}+dx^q+ax+a^2+ad+d^2=(x+a\omega+d\omega^2)(x+a\omega^2+d\omega)\prod\limits_{k\in T_2}g_k(x).$$

\end{thm}
\begin{proof}  Let $\alpha$ be a root of a polynomial $g_k(x)=x^3+kx^2+(a^2+k(a+d)+ad+d^2)x+a^3+kad+d^3$ over $\Bbb F_q$, then $$g_k(x)=(x-\alpha)(x-A(\alpha))(x-A^2(\alpha))$$ and
 $$k=\alpha+A(\alpha)+A^2(\alpha)=\frac{\alpha^3+(a^2+d^2)\alpha+a^3+d^3}{(a+\alpha)(d+\alpha)}\in \mathbb{F}_{q}.$$ In the following, we use Cardano's method to find  the condition of  $k\in \Bbb F_q$ such that $g_k(x)$ is irreducible over $\Bbb F_q$.

Now we make some  variable substitutions for $g_k(x)$.
Let $x=y+k$, then  $g_k(x)=y^3+(k^2+a^2+d^2+ad+k(a+d))y+k(a^2+d^2)+k^2(a+d)+a^3+d^3$.

If $k^2+a^2+d^2+ad+k(a+d)=0$, then $g_k(x)=(x+k)^3$, $x=k$ is a root of $g_k(x)$, and $A(\alpha)=\alpha$.

If $k^2+a^2+d^2+ad+k(a+d)\neq0$, i.e., $A(\alpha)\ne \alpha$, then
 by $y=z(k^2+a^2+d^2+ad+k(a+d))^{\frac{1}{2}}$,    $g_k(x)=(k^2+a^2+d^2+ad+k(a+d))^{\frac{3}{2}}(z^3+z+t)$,  where $$t=\frac{a+d}{(k^2+a^2+d^2+ad+k(a+d))^{\frac{1}{2}}}.$$
 Set   $z=\mu+\nu$ and  $(\mu+\nu)^3+(\mu+\nu)+t=0$.

 Suppose that
$$\mu\nu=1 \mathrm{~and~} \mu^3+\nu^3=t.$$
 Then by $\mu^3+\frac{1}{\mu^3}=t$,
 $$\frac 1{\mu^6+1}+\frac 1{(\mu^6+1)^2}= \frac{\mu^6}{(\mu^6+1)^2}=(\frac{1}{t})^2=(\frac{k+a}{a+d})^2+\frac{k+a}{a+d}+1.$$
Hence $$(\frac{1}{\mu^6+1}+\frac{k+a}{a+d})^2+(\frac{1}{\mu^6+1}+\frac{k+a}{a+d})+1=0$$ and $$\frac{1}{\mu^6+1}+\frac{k+a}{a+d}=\omega^i, i=1,2,$$
where $\omega\in\Bbb F_q$ and $\ord(\omega)=3$.
So $$\mu^6=\frac{a+d}{k+a\omega+d\omega^2}+1 \mbox{ or }\mu^6=\frac{a+d}{k+a\omega^2+d\omega}+1.$$

Without loss of generality, we assume that
 $\mu^6=\frac{a+d}{k+a\omega+d\omega^2}+1$ and $\nu^6=\frac{a+d}{k+a\omega^2+d\omega}+1$.

If $\mu^6=\frac{a+d}{k+a\omega+d\omega^2}+1=c\in D_0$ ( $c\ne 1$) and $k\in T_0=\{\frac{a+d}{c+1}+a\omega+d\omega^2: c\in D_0\}$, then $u, v\in \Bbb F_q$ and  $z^3+z+t$ has three roots in $\mathbb{F}_{q}$:
 $z_1=\mu+\nu$, $z_2=\omega\mu+\omega^2\nu$, $z_3=\omega^2\mu+\omega\nu$. Hence $g_k(x)=x^3+kx^2+(a^2+k(a+d)+ad+d^2)x+a^3+kad+d^3$ has three roots in $\Bbb F_q$:
   \begin{eqnarray}
   \alpha&=&\frac{a+d}{\mu^2+1}+a\omega+d\omega^2, A(\alpha)=\frac{a+d}{\omega^2\mu^2+1}+a\omega+d\omega^2,\\ &&A^2(\alpha)=\frac{a+d}{\omega\mu^2+1}+a\omega+d\omega^2.\nonumber
   \end{eqnarray}

If $\mu^6=\frac{a+d}{k+a\omega+d\omega^2}+1=c\in D_1$ and $k\in T_1=\{\frac{a+d}{c+1}+a\omega+d\omega^2: c\in D_1\}$, then $\mu,\nu \in\mathbb{F}_{q^3}\backslash\mathbb{F}_{q}$ and  $g_k(x)=x^3+kx^2+(a^2+k(a+d)+ad+d^2)x+a^3+kad+d^3$ has three roots: $\alpha, A(\alpha), A^2(\alpha)\in
\mathbb{F}_{q^3}\backslash\mathbb{F}_{q}$ as (3.5). Since $\beta$ is a root of $x^3-\xi$ such that $\beta^q=\omega \beta$, $(\mu^2)^q=\omega\mu^2$ by $\mu^6\in D_1$.
Hence  $q$ is the least positive integer such that
\begin{eqnarray*}\alpha^q=(\frac{a+d}{\mu^2+1}+a\omega+d\omega^2)^q =\frac{a+d}{\omega\mu^2+1}+a\omega+d\omega^2=A^2(\alpha).\end{eqnarray*}

Moreover, if $\alpha$ is a root of the irreducible polynomial $g_k(x)$ for $k\in T_1$, then $A^2(\alpha)=\alpha^q$ and $\alpha$ is a root of $h(x)=x^{q+1}+ax^q+dx+a^2+ad+d^2$, so $g_k(x)|h(x)$. It is clear that $a\omega+d\omega^2$ and $a\omega^2+d\omega$  are also  roots of $h(x)$.
If $k_1\ne k_2\in T_1$, then $\gcd(g_{k_1}(x),g_{k_2}(x))=1$ and $|T_1|=\frac{q-1}3$.
Hence
$$h(x)=x^{q+1}+ax^q+dx+a^2+ad+d^2=(x+a\omega+d\omega^2)(x+a\omega^2+d\omega)\prod\limits_{k\in T_1}g_k(x).$$

If $\mu^6=\frac{a+d}{k+a\omega+d\omega^2}+1=c\in D_2$  and $k\in T_2=\{\frac{a+d}{c+1}+a\omega+d\omega^2: c\in D_2\}$, then $\mu,\nu \in\mathbb{F}_{q^3}\backslash\mathbb{F}_{q}$,  $g_k(x)=x^3+kx^2+(a^2+k(a+d)+ad+d^2)x+a^3+kad+d^3$ has three roots: $\alpha, A(\alpha), A^2(\alpha)\in
\mathbb{F}_{q^3}\backslash\mathbb{F}_{q}$, and  $(\mu^2)^q=\omega^2\mu^2$.
Similarly,   $q$ is the least positive integer such that
$$\alpha^q=(\frac{a+d}{\mu^2+1}+a\omega+d\omega^2)^q=\frac{a+d}{\omega^2\mu^2+1}+a\omega+d\omega^2=A(\alpha),$$
$$h(x)=x^{q+1}+dx^q+ax+a^2+ad+d^2=(x+a\omega+dw^2)(x+a\omega^2+d\omega)\prod\limits_{k\in T_2}g_k(x).$$

The proof of this theorem is done.
\end{proof}

Now we consider Theorem 3.3 in the case $r=3s$.

\begin{thm}{\label{th9}} Let $3\mid(q-1)$, $\xi$ be a primitive element of $\mathbb{F}_{q^s}$ and  $\omega=\xi^{\frac{q^s-1}{3}}$ be an element of order $3$. Let
 $\beta$ be a root of the irreducible polynomial $x^3-\xi$ over $\Bbb F_{q^s}$ such that $\beta^{q^s}=\omega\beta$. Let $A=\left(\begin{array}{cc} a&a^2+ad+d^2\\1&d\end{array}\right)\in PGL_2(\mathbb{F}_{q})$ be a matrix of order $3$.

(1) Suppose that $s=t^e$, where $t $ is an odd prime and $e$ is a positive integer. Then for $0\leq i\leq e$,

\begin{eqnarray*}g_{i,k}(x)&=&\prod^{t^i-1}\limits_{j=0}(x^3+k^{q^j}x^2+(a^2+k^{q^j}(a+d)+ad+d^2)x+a^3+k^{q^j}ad+d^3)\\ &=&\prod^{t^i-1}\limits_{j=0}(x-\alpha^{q^j})(x-A(\alpha^{q^j}))(x-A^2(\alpha^{q^j})),\alpha\in\Omega\end{eqnarray*}
is irreducible over $\mathbb{F}_{q}$ if and only if $k\in (T^{(i)}_1\cup T^{(i)}_2)\cap(\Bbb F_{q^{t^i}}\backslash\Bbb F_{q^{t^{i-1}}})$, where $T_j^{(i)}=\{\frac{a+d}{c+1}+a\omega+d\omega^2: c\in D^{(i)}_j\}$, $\xi_i=\xi^{\frac{q^s-1}{q^{t^i}-1}}$, $D^{(i)}_j=\xi_i^j<\xi^3_i>$,  $j=1,2$.  In fact, $q^{t^i}$ is the least positive integer such that $A^2(\alpha)=\alpha^{q^{t^i}}$ for $k\in T^{(i)}_1$; $q^{t^i}$ is the least positive integer such that $A(\alpha)=\alpha^{q^{t^i}}$ for $k\in T^{(i)}_2$.

 Moreover, there are two irreducible factorizations of $h(x)$ over $\Bbb F_q$:
 \begin{eqnarray*}h(x)&=&x^{q^s+1}+ax^{q^s}+dx+a^2+ad+d^2\\ &=&(x+a\omega+d\omega^2)(x+a\omega^2+d\omega)\prod_{i=0}^e\prod\limits_{k\in (T^{(i)}_1\cap(\Bbb F_{q^{t^i}}\backslash\Bbb F_{q^{t^{i-1}}}))/\sim}g_{i,k}(x),\end{eqnarray*}
 where  $({T_1^{(i)}}\cap(\Bbb F_{q^{t^i}}\backslash\Bbb F_{q^{t^{i-1}}}))/\thicksim$ is a quotient set, and  $\sim$ is an equivalent relation of a set ${T_1^{(i)}}\cap(\Bbb F_{q^{t^i}}\backslash\Bbb F_{q^{t^{i-1}}})$: for $k', k\in {T_1^{(i)}}\cap(\Bbb F_{q^{t^i}}\backslash\Bbb F_{q^{t^{i-1}}})$, $k'\sim k$ if and only if $k'=k^{q^j}, 0\le j\le t^i-1$.

  And
 \begin{eqnarray*}h(x)&=&x^{q^s+1}+dx^{q^s}+ax+a^2+ad+d^2\\ &=&(x+a\omega+d\omega^2)(x+a\omega^2+d\omega)\prod_{i=0}^e\prod\limits_{k\in (T^{(i)}_2\cap(\Bbb F_{q^{t^i}}\backslash\Bbb F_{q^{t^{i-1}}}))/\sim }g_{i,k}(x),\end{eqnarray*}
where all $\sim$ are equivalent relations as above.

(2) Suppose that  $s=t_1t_2$, where $t_i>3(i=1,2)$ are two distinct primes. Then
$$g_k(x)=x^3+kx^2+(a^2+k(a+d)+ad+d^2)x+a^3+kad+d^3=\prod^{2}\limits_{i=0}(x-A^i(\alpha)), \alpha\in\Omega$$ is irreducible over $\mathbb{F}_{q}$ if and only if $k\in T^{(0)}_1\cup T^{(0)}_2$, where $T_j^{(0)}=\{\frac{a+d}{c+1}+a\omega+d\omega^2: c\in D^{(0)}_j\}$, $\xi_0=\xi^{\frac{q^s-1}{q-1}}$, $D^{(0)}_j=\xi_0^j<\xi^3_0>$,  $j=1,2$. In fact, $q$ is the least positive integer such that $A^2(\alpha)=\alpha^{q}$ for $k\in T^{(0)}_1$; $q$ is the least positive integer such that $A(\alpha)=\alpha^{q}$ for  $k\in T^{(0)}_2$.

For $t_i(i=1,2)$,
\begin{eqnarray*}g_{t_{i}, k}(x)&=&\prod^{t_i-1}\limits_{j=0}
(x^3+k^{q^j}x^2+(a^2+k^{q^j}(a+d)+ad+d^2)x+a^3+k^{q^j}ad+d^3)\\ &=&\prod_{j=0}^{t_i-1}(x-\alpha^{q^j})(x-A(\alpha^{q^j}))(x-A^2(\alpha^{q^j})),\alpha\in\Omega
\end{eqnarray*}
is  irreducible over $\Bbb F_q$
 if and only if  $k\in (T^{(t_i)}_1\cup T^{(t_i)}_2)\cap(\Bbb F_{q^{t_i}}\backslash\Bbb F_{q})$, where $T^{(t_i)}_j=\{\frac{a+d}{c+1}+a\omega+d\omega^2: c\in D^{(t_i)}_{j}\}$, $\xi_{t_i}=\xi^{\frac{q^s-1}{q^{t_i}-1}}$, $D^{(t_i)}_{j}=\xi^j_{t_i}<\xi^3_{t_i}>$, $j=1,2$. In fact,  $q^{t_i}$ is the least positive integer such that  $A^2(\alpha)=\alpha^{q^{t_i}}$ for $k\in T^{(t_i)}_1$;  $q^{t_i}$ is the least positive integer such that  $A(\alpha)=\alpha^{q^{t_i}}$ for $k\in T^{(t_i)}_2$.

For $s=t_1t_2$,
  \begin{eqnarray*}g_{s, k}(x)&=&\prod^{s-1}\limits_{j=0}(x^3+k^{q^j}x^2+(a^2+k^{q^j}
  (a+d)+ad+d^2)x+a^3+k^{q^j}ad+d^3)\\ &=&\prod^{s-1}\limits_{j=0}(x-\alpha^{q^j})(x-A(\alpha^{q^j}))(x-A^2(\alpha^{q^j})), \alpha\in\Omega\end{eqnarray*}  is irreducible over $\Bbb F_q$
 if and only if
  $k\in({T^{(s)}_1}\cup{T^{(s)}_2})\cap(\Bbb F_{q^s}\backslash(\Bbb F_{q^{t_1}}\cup\Bbb F_{q^{t_2}}))$, where $T^{(s)}_{j}=\{\frac{a+d}{c+1}+a\omega+d\omega^2: c\in D^{(s)}_j\}$,  $D^{(s)}_j=\xi^j<\xi^3>, j=1,2$. In fact, $q^s$ is the least positive integer such that  $A^2(\alpha)=\alpha^{q^{s}}$ for $k\in T^{(s)}_{1}$; $q^s$ is the least positive integer such that  $A(\alpha)=\alpha^{q^{s}}$ for $k\in T^{(s)}_{2}$.

  Moreover, there are two irreducible factorizations of $h(x)$ over $\Bbb F_q$:
 \begin{eqnarray}h(x)&=&x^{q^s+1}+ax^{q^s}+dx+a^2+ad+d^2\nonumber\\
 &=&(x+a\omega+d\omega^2)(x+a\omega^2+d\omega)\prod\limits_{k\in T^{(0)}_1}g_{k}(x)\prod\limits_{k\in({T^{(t_1)}_1}\cap(\Bbb F_{q^{t_1}}\backslash\Bbb F_{q}))/\sim}g_{t_1,k}(x)\nonumber\\
 &&\prod\limits_{k\in({T^{(t_2)}_1}\cap(\Bbb F_{q^{t_2}}\backslash\Bbb F_{q}))/\sim}g_{t_2,k}(x)
 \prod\limits_{k\in{(T^{(s)}_1}\cap(\Bbb F_{q^s}\backslash(\Bbb F_{q^{t_1}}\cup\Bbb F_{q^{t_2}})))/\sim}g_{s,k}(x),
 \end{eqnarray}
and
  \begin{eqnarray}h(x)&=&x^{q^s+1}+dx^{q^s}+ax+a^2+ad+d^2\nonumber\\
 &=&(x+a\omega+d\omega^2)(x+a\omega^2+d\omega)\prod\limits_{k\in T^{(0)}_2}g_{k}(x)\prod\limits_{k\in({T^{(t_1)}_2}\cap(\Bbb F_{q^{t_1}}\backslash\Bbb F_{q}))/\sim}g_{t_1,k}(x)\nonumber\\
 &&\prod\limits_{k\in({T^{(t_2)}_2}\cap(\Bbb F_{q^{t_2}}\backslash\Bbb F_{q}))/\sim}g_{t_2,k}(x)
 \prod\limits_{k\in{(T^{(s)}_2}\cap(\Bbb F_{q^s}\backslash(\Bbb F_{q^{t_1}}\cup\Bbb F_{q^{t_2}})))/\sim}g_{s,k}(x),
 \end{eqnarray}
 where all $\sim$ are equivalent relations as above.
\end{thm}
\begin{proof}
(1) Now we consider the case:  $s=t^e$, where $t $ is an odd prime and $e$ is a positive integer. Let $\xi_i=\xi^{\frac{q^s-1}{q^{t^i}-1}}$ be a primitive element of $\mathbb{F}_{q^{t^i}}$, where $0\leq i\leq e$. Since  $\beta$ is a root of the irreducible polynomial $x^3-\xi$ over $\Bbb F_{q^s}$ such that $\beta^{q^s}=\omega\beta$, $\beta_i$ is a root of the irreducible polynomial $x^3-\xi_i$ over $\Bbb F_{q^{t^i}}$, where $\beta_i=\beta^{\frac{q^s-1}{q^{t^i}-1}}$.
  Then $$\beta^{q^{t^i}}_i=\beta_i\beta_i^{q^{t^i}-1}=\beta_i\beta^{q^s-1}=\omega\beta_i.$$ By the proofs of Theorems \ref{th7}(1) and 3.8,  the proof of (1) is done.

(2) Next we consider the case:   $s=t_1t_2$, where $t_i>3(i=1,2)$ are two distinct primes.
  Let $\xi_{0}=\xi^{\frac{q^s-1}{q-1}}$ be a primitive element of $\mathbb{F}_{q}$ and $\xi_{t_i}=\xi^{\frac{q^s-1}{q^{t_i}-1}}$ be a primitive element of $\mathbb{F}_{q^{t_i}}$, where $i=1,2$. Since  $\beta$ is a root of the irreducible polynomial $x^3-\xi$ over $\Bbb F_{q^s}$ such that $\beta^{q^s}=\omega\beta$, $\beta_0$ is a root of the irreducible polynomial $x^3-\xi_0$ over $\Bbb F_{q}$, where $\beta_{0}=\beta^{\frac{q^s-1}{q-1}}$ and $\beta_{t_i}$ is a root of the irreducible polynomial $x^3-\xi_{t_i}$ over $\mathbb{F}_{q^{t_i}}$, where $\beta_{t_i}=\beta^{\frac{q^s-1}{q^{t_i}-1}}$. Then
   $$\beta^q_{0}=\beta_0\beta_0^{q-1}=\beta_0\beta^{q^s-1}=\omega\beta_0$$ and $${\beta^{q^{t_i}}_{t_i}}=\beta_{t_i}\beta_{t_i}^{q^{t_i}-1}=\beta_{t_i}\beta^{q^s-1}=\omega\beta_{t_i}.$$
   By the proofs of Theorems \ref{th7} (2) and 3.8,  the proof of (2)  is done.
\end{proof}
All the examples are obtained using the MAGMA system.
\begin{exa}Let $\xi$ be a root of an  irreducible polynomial $x^3+x+1$ over $\Bbb F_2$ and $\Bbb F_{8}=\Bbb F_2(\xi)$.
 Let $A=\left(\begin{array}{cc} 1&0\\1&1\end{array}\right)\in PGL_2(\mathbb{F}_{8})$. Following Theorem \ref{th6}, if $k\in\{1, \xi, \xi^2, \xi^{4}\}$, then the irreducible polynomials $g(x)$ over $\mathbb{F}_{8}$ as follows: $$g_1(x)=x^2 + \xi x +\xi;$$ $$g_2(x)=x^2 + \xi^2 x + \xi^2;$$ $$g_3(x)=x^2 + \xi^{4}x + \xi^{4};$$ $$g_4(x)=x^2 +x +1.$$
\end{exa}

\begin{exa}
Let $s=5$ and $\xi$ be a primitive element of $\Bbb F_{32}$. Let ${A}=\left(\begin{array}{cc} 1&0\\1&1\end{array}\right)\in PGL_2(\mathbb{F}_{2})$. Following Theorem \ref{th7}, if $$k\in\{\xi^5, \xi^7, \xi^{9}, \xi^{10}, \xi^{11}, \xi^{13}, \xi^{14}, \xi^{18}, \xi^{19},  \xi^{20}, \xi^{21}, \xi^{22}, \xi^{25}, \xi^{26}, \xi^{28}\},$$ then the irreducible polynomials $g(x)$ over $\mathbb{F}_{2}$ as follows: $$g_1(x)=x^{10} + x^8 + x^7 + x^6 + x^2 + x + 1;$$ $$g_2(x)=x^{10} + x^9 + x^8 + x^7 + x^2 + x + 1;$$ $$g_3(x)=x^{10} + x^9 + x^5 + x^4 + x^2 + x + 1.$$
\end{exa}
\begin{exa}Let $\xi$ be a root of an  irreducible polynomial $x^4+x+1$ over $\Bbb F_2$ and $\Bbb F_{16}=\Bbb F_2(\xi)$.
  Let $A=\left(\begin{array}{cc} 1&0\\1&\xi^{5}\end{array}\right)\in PGL_2(\mathbb{F}_{16})$. Following Theorem \ref{th8}, if $k\in \{\xi^9, \xi^6, \xi^5, \xi, \xi^4\}$,   then $A(\alpha)=\alpha^{16}$ and the irreducible polynomials $g(x)$ over $\mathbb{F}_{16}$ as follows: $$g_1(x)=x^3 + \xi^9x^2 + \xi^4x +\xi^{14};$$ $$g_2(x)=x^3 +\xi^6x^2 + \xi x + \xi^{11};$$ $$g_3(x)=x^3 + \xi^5x^2 + x + \xi^{10};$$ $$g_4(x)=x^3 + \xi x^2 + \xi^{11}x + \xi^6;$$  $$g_5(x)=x^3 + \xi^4x^2 + \xi^{14}x + \xi^9.$$
If $k\in\{ \xi^8, \xi^7, 1, \xi^2, \xi^{13}\}$,  then ${A}^2(\alpha)=\alpha^{16}$ and the irreducible polynomials $g(x)$ over $\mathbb{F}_{16}$ as follows: $$g_1(x)=x^3 + \xi^8x^2 + \xi^3x +\xi^{13};$$ $$g_2(x)=x^3 +\xi^7x^2 + \xi^2 x + \xi^{12};$$ $$g_3(x)=x^3 + x^2 +\xi^{10}x + \xi^{5};$$ $$g_4(x)=x^3 + \xi^2 x^2 + \xi^{12}x + \xi^7;$$  $$g_5(x)=x^3 + \xi^{13}x^2 + \xi^{8}x + \xi^3.$$
\end{exa}

\section {Binary irreducible quasi-cyclic parity-check subcodes of Goppa codes and extended Goppa codes}
It is well-known that  extended  Goppa codes and parity check
subcodes of Goppa codes are  Alternant codes associated to a Goppa polynomial $g(x)$ ( see  Section 2.1).
In this section, we will study binary irreducible quasi-cyclic
Alternant codes  $\mathcal{A}_{r+1}(v_{g,L},L)$ with  Goppa polynomials $g(x)$.

 \subsection{Permutation automorphisms of  Alternant codes}
\begin{defn}
 Let   $\mathcal{C}$ be a code of length $m$ and $\psi$  a  permutation of order $m$, we define:
 $$\mathcal{C}^\psi=\{\mathbf{c}^\psi=(c_{\psi(0)},c_{\psi(1)},\ldots, c_{\psi(m-1)})|\mathbf{c}=(c_0,c_1,\ldots, c_{m-1})\in\mathcal{C}\}.$$
 If $\mathbf c\in\mathcal C$, then $\mathbf c^{\psi}\in \mathcal C$. The permutation $\psi$ is called  a {\it   permutation automorphism} of $\mathcal{C}$.
\end{defn}
\begin{lem}\label{le1}{\cite{6}}
Let $g(x)$ be a polynomial of degree $r$ over $\mathbb{F}_{q}$ and $L=(\alpha_0,\ldots,\alpha_{m-1})$ an ordered tuples of $m$ distinct points in the projective line set $\overline{\mathbb{F}}_{q}$. Let $A=\left(\begin{array}{cc} a&b\\1&d\end{array}\right)\in PGL_2(\Bbb F_{q})$, $L'=(A^{-1}(\alpha_0),\ldots,A^{-1}(\alpha_{m-1}))$, and $g'(x)=(x+d)^rg(A(x))$, where $A^{-1}(x)= \frac{dx+b}{x+a}$
and $g(a)\ne 0$. Then the Alternant code $\mathcal{A}_{r+1}(v_{g,L},L)$ is equal to the Alternant code  $\mathcal{A}_{r+1}(v_{g',L'},L')$.
\end{lem}
The alternant  codes that will be obtained here correspond
to permutation groups by the action of projective linear maps
$x \rightarrow \frac{dx+b}{x+a}$ on the support $(\alpha_0, \alpha_1,\ldots, \alpha_{m-1})$ of the alternant code. If this support is globally
invariant by this projective linear map, then this
induces a permutation $\psi$ of the code positions $\{0, 1,\cdots, m-1\}$
by defining $\psi(i)$ as the unique integer in $\{0, 1,\cdots, m-1\}$
such that $\alpha_{\psi (i)} = \frac{d\alpha_i+b}{\alpha_i+a}$. In such a case, we say that  $\psi$ is the
permutation induced by the projective linear map $x \rightarrow \frac{dx+b}{x+a}$.

  By
\cite[Theorem 1]{2} and \ref{le1}, there is the following result.
\begin{prop}\label{c1}
Let $A=\left(\begin{array}{ll} a&b\\1&d\end{array}\right)\in PGL_2(\Bbb F_q)$ and
$L=(\alpha_0,\ldots,\alpha_{m-1})$  be a support
which is globally invariant by the
projective linear map $x \rightarrow A^{-1}(x)=\frac{dx+b}{x+a}$. Let $\psi$ be the permutation
induced by this projective linear map.  Assume
that  $g(x)$ is a polynomial of degree $r$ over $\mathbb{F}_{q}$  such that
 $\gamma g(x)=(x+d)^rg(A(x))$,
  where $g({a})\ne 0$ and $0\ne \gamma\in \Bbb F_q$. Then $\psi$ is a permutation automorphism of the alternant code $\mathcal{A}_{r+1}(v_{g,L},L)$, i.e., ${\mathbf c}^{\psi}=(c_{\psi (0)}, c_{\psi(1)},\ldots, c_{\psi(m-1)})\in \mathcal{A}_{r+1}(v_{g,L},L)$ if
   $\mathbf c=(c_0,c_1, \ldots, c_{m-1})\in \mathcal{A}_{r+1}(v_{g,L},L)$, where $A^{-1}(\alpha_i)=\frac{d\alpha_i+b}{\alpha_i+a}=\alpha_{\psi(i)}$.
\end{prop}

\subsection {Binary  irreducible
quasi-cyclic Alternant codes}
\begin{defn}
 Let $\mathcal{C}$ be an alternant code of length $m$ defined
over a finite field $\mathbb{F}_{q}$ and $G$ the permutation group of $\mathcal C$.
Given a nonnegative integer $\lambda \leq m$, we say that $\mathcal{C}$ is Quasi-Cyclic (QC) if $G$ contains a subgroup of the form
$(Z/\lambda Z)$.
\end{defn}
\begin{thm}{\label{th3}}
Let $\langle A \rangle $ be the cyclic group acting  on $\overline {\Bbb F}_q$, where $A=\left(\begin{array}{cc} a&b\\1&d\end{array}\right)\in PGL_2(\Bbb F_{q})$ is of order $l$ and $l$ is prime.
Let $L=(L_1,\ldots, L_{\tau})$ be a support, where $L_1, \ldots, L_{\tau}$ are   some disjoint orbits  and $|L_i|=l, i=1,\ldots, \tau$.  Let $g(x)=\prod^{s-1}\limits_{j=0}\prod^{l-1}\limits_{i=0}(x-A^i(\beta^{q^j}))$ be the Goppa polynomial of degree $r=sl$ over $\Bbb F_{q}$  and  $s$  be the least positive integer such that $A^u(\beta)=\beta^{q^s}$, $1\le u\le l-1$.  Then $\mathcal{A}_{r+1}(v_{g,L},L)$  is  a binary irreducible quasi-cyclic  Alternant code over $\Bbb F_{q}$.
\end{thm}
\begin{proof}
By Theorem \ref{th1}, $g(x)$ is an irreducible polynomial over $\Bbb F_{q}$. By Lemma \ref{c11},  $\gamma g(x)=(x+d)^rg({A}(x))$, where $0\ne \gamma\in \Bbb F_q$. Since  $L_1, \ldots, L_{\tau}$   are some disjoint orbits, $L$ is globally invariant by the
projective linear map $x \rightarrow A^{-1}(x)=\frac{dx+b}{x+a}$.  In summary, $\mathcal{A}_{r+1}(v_{g,L},L)$  is a binary irreducible quasi-cyclic Alternant code over $\Bbb F_{q}$. The proof of this theorem is done.
\end{proof}

All the examples are obtained using the MAGMA system.

\begin{exa}
 Let $\xi$ be a root of an irreducible polynomial $x^6 + x^4 + x^3 + x + 1$ over $\Bbb F_2$ and $\Bbb F_{64}=\Bbb F_2(\xi)$.  Let $A=\left(\begin{array}{cc} 1&0\\1&\xi^{21}\end{array}\right)\in PGL_2(\mathbb{F}_{64})$.  The orbits of $\overline{\mathbb{F}}_{64}$ are: $L_1=(0)$; $L_2=(\xi, \xi^{6}, \xi^{29})$;
$L_3=(\xi^2, \xi^{15}, \xi^{37})$; $L_4=(\xi^3, \xi^{9}, \xi^{11})$; $L_5=(\xi^4, \xi^{24}, \xi^{53})$; $L_6=(\xi^5, \xi^{49}, \xi^{59})$; $L_7=(\xi^{7}, \xi^{47}, \xi^{20})$; $L_8=(\xi^{8}, \xi^{60}, \xi^{22}); L_{9}=(\xi^{10}, \xi^{40}, \xi^{34});$ $L_{10}=(\xi^{12}, \xi^{36}, \xi^{44});$ $L_{11}=(\xi^{13}, \xi^{56}, \xi^{31});$ $L_{12}=(\xi^{14}, \xi^{55}, \xi^{19});$ $L_{13}=(\xi^{16}, \xi^{33}, \xi^{23});$ $L_{14}=(\xi^{17}, \xi^{28}, \xi^{62});$ $L_{15}=(\xi^{18}, \xi^{50}, \xi^{48});$ $L_{16}=(\xi^{25}, \xi^{32}, \xi^{51});$ $L_{17}=(\xi^{26}, \xi^{38}, \xi^{41});$ $L_{18}=(\xi^{27}, \xi^{43}, \xi^{39});$ $L_{19}=(\xi^{30}, \xi^{45}, \xi^{46} );$ $L_{20}=(\xi^{35},\xi^{61}, \xi^{52});$ $L_{21}=(\xi^{54},\xi^{58}, \xi^{57}); L_{22}=(\xi^{21},\infty, 1); L_{23}=(\xi^{42})$.

 We compute the  polynomials of degree $3$ satisfying  Theorem \ref{th3}, $g(\alpha)=0$  and ${A}(\alpha)=\alpha^{q}$ as follows:
$$g_1(x)=x^3 + \xi^{28}x^2 + \xi^7x +\xi^{49};$$ $$g_2(x)=x^3 +\xi^{17}x^2 + \xi^{59} x + \xi^{38};$$ $$ g_3(x)=x^3 + \xi^{49}x^2 + \xi^{28}x + \xi^{7};$$ $$ g_4(x)=x^3 + \xi^{43} x^2 + \xi^{22}x + \xi;$$ $$ g_5(x)=x^3 + \xi^{59}x^2 + \xi^{38}x + \xi^{17};$$ $$ g_6(x)=x^3 + \xi^{5}x^2 + \xi^{47}x + \xi^{26};$$ $$g_7(x)=x^3 + \xi^{27}x^2 + \xi^{6}x + \xi^{48};$$ $$ g_8(x)=x^3 + \xi^{7}x^2 + \xi^{49}x + \xi^{28};$$ $$ g_9(x)=x^3 + \xi^{62}x^2 + \xi^{41}x + \xi^{20};$$ $$g_{10}(x)=x^3 + \xi^{46}x^2 + \xi^{25}x + \xi^{4};$$ $$ g_{11}(x)=x^3 + \xi^{39}x^2 + \xi^{18}x + \xi^{60};$$ $$ g_{12}(x)=x^3 + \xi^{30}x^2 + \xi^{9}x + \xi^{51};$$ $$ g_{13}(x)=x^3 + \xi^{10}x^2 + \xi^{52}x + \xi^{31};$$ $$ g_{14}(x)=x^3 + \xi^{54}x^2 + \xi^{33}x + \xi^{12};$$ $$ g_{15}(x)=x^3 + \xi^{47}x^2 + \xi^{26}x + \xi^{5};$$ $$ g_{16}(x)=x^3 + \xi^{58}x^2 + \xi^{37}x + \xi^{16};$$ $$ g_{17}(x)=x^3 + \xi^{40}x^2 + \xi^{19}x + \xi^{61};$$ $$ g_{18}(x)=x^3 + \xi^{57}x^2 + \xi^{36}x + \xi^{15};$$ $$ g_{19}(x)=x^3 + \xi^{34}x^2 + \xi^{13}x + \xi^{55};$$ $$ g_{20}(x)=x^3 + \xi^{45}x^2 + \xi^{24}x + \xi^{3};$$ $$ g_{21}(x)=x^3 + \xi^{20}x^2 + \xi^{62}x + \xi^{41}.$$
  Let $g_i(x)$  be the Goppa polynomial of  Alternant code $\mathcal{A}_{4}(v_{g,L},L)$, where $1\leq i\leq 21$. If $L=(L_2,\ldots,L_{22})$, then $\mathcal{A}_{4}(v_{g,L},L)$ is a binary irreducible $21$-quasi-cyclic extended Goppa code with length $m=63$; If $L=(L_2,\ldots,L_{21})$, then $\mathcal{A}_{4}(v_{g,L},L)$ is a binary irreducible $20$-quasi-cyclic parity-check subcode of Goppa code with length $m=60$.
\end{exa}\begin{exa}
 Let $\xi$ be a root of an irreducible polynomial $x^6 + x^4 + x^3 + x + 1$ over $\Bbb F_2$ and $\Bbb F_{64}=\Bbb F_2(\xi)$.  Let ${A}=\left(\begin{array}{cc} \xi^9&0\\1&1\end{array}\right)\in PGL_2(\mathbb{F}_{64})$. The orbits of $\overline{\mathbb{F}}_{64}$ are:
$L_1=(0); L_2=(\xi, \xi^{17}, \xi^{50}, \xi^{6}, \xi^{52}, \xi^{49}, \xi^{56}); L_3=(\xi^2, \xi^{25}, \xi^{39}, \xi^{31}, \xi^{44}, \xi^{24},\xi^{55} ); L_4=(\xi^3, \xi^{62}, \xi^{16}, \xi^{11}, \xi^{60}, \xi^{59}, \xi^{37}); L_5=(\xi^4, \xi^{41}, \xi^{26}, \xi^{29},$ $ \xi^{57}, \xi^{46}, \xi^{33}); L_6=(\xi^5, \xi^{47}, \xi^{58}, \xi^{42}, $ $\xi^{30}, \xi^{34}, \xi^{28}); L_7=(\xi^{7}, \xi^{8}, \xi^{10}, \xi^{22}, \xi^{48}, \xi^{38}, \xi^{14}); L_8=(\xi^{12}, \xi^{32}, \xi^{13},\xi^{19},\xi^{43}, \xi^{15}, \xi^{53}); L_{9}=(\xi^{20}, \xi^{35}, \xi^{40}, \xi^{61}, \xi^{23}, \xi^{21}, \xi^{51}); L_{10}=(\xi^{9}, \xi^{54}, \xi^{45}, \xi^{18}, $ $ \xi^{36}, 1, \infty); L_{11}=(\xi^{27}).$

 We compute the  polynomials of degree $7$ satisfying  Theorem \ref{th3}, $g(\alpha)=0$ and ${A}(\alpha)=\alpha^{q}$ as follows:
$$ g_1(x)=x^7 +\xi^{35}x^6 + \xi^{62}x^5 + \xi^{26}x^4 + \xi^{53}x^3 + \xi^{17}x^2 + \xi^{44}x + \xi^8;$$
$$ g_2(x)=x^7 + \xi x^6 + \xi^{28}x^5 + \xi^{55}x^4 + \xi^{19}x^3 + \xi^{46}x^2 + \xi^{10}x + \xi^{37};$$
$$ g_3(x)=x^7 + \xi^{28}x^6 + \xi^{55}x^5 + \xi^{19}x^4 + \xi^{46}x^3 + \xi^{10}x^2 + \xi^{37}x + \xi;$$
$$ g_4(x)=x^7 + \xi^{18}x^6 +\xi^{45}x^5 + \xi^9x^4 + \xi^{36}x^3 + x^2 + \xi^{27}x + \xi^{54};$$
$$ g_5(x)=x^7 +\xi^{24}x^6 + \xi^{51}x^5 +\xi^{15}x^4 + \xi^{42}x^3 + \xi^6x^2 + \xi^{33}x + \xi^{60};$$
$$ g_6(x)=x^7 + \xi^3x^6 + \xi^{30}x^5 + \xi^{57}x^4 + \xi^{21}x^3 + \xi^{48}x^2 + \xi^{12}x + \xi^{39};$$
$$ g_7(x)=x^7 + \xi^{12}x^6 + \xi^{39}x^5 + \xi^3x^4 + \xi^{30}x^3 + \xi^{57}x^2 + \xi^{21}x + \xi^{48};$$
$$ g_8(x)=x^7 + \xi^{33}x^6 + \xi^{60}x^5 +\xi^{24}x^4 + \xi^{51}x^3 + \xi^{15}x^2 + \xi^{42}x + \xi^6;$$
$$ g_9(x)=x^7 + \xi^8x^6 + \xi^{35}x^5 + \xi^{62}x^4 + \xi^{26}x^3 + \xi^{53}x^2 + \xi^{17}x + \xi^{44}.$$
 Let $g_i(x)$ be the Goppa polynomial of  Alternant code $\mathcal{A}_{8}(v_{g,L},L)$, where $1\leq i\leq 9$. If $L=(L_2,\ldots,L_{10})$, then $\mathcal{A}_{8}(v_{g,L},L)$ is a binary irreducible $9$-quasi-cyclic extended Goppa code with length $m=63$; If $L=(L_2,\ldots,L_9)$, then $\mathcal{A}_{8}(v_{g,L},L)$ is a binary irreducible $8$-quasi-cyclic parity-check subcode of Goppa code with length $m=56$.
\end{exa}
In the following,  we shall give specific constructions of binary irreducible quasi-cyclic parity-check subcodes  of Goppa codes over $\mathbb{F}_{{q}^{2}}$.

\begin{cor}{\label{c12}}
Let $\langle A \rangle $ be the cyclic group acting  on $U_{q+1}$, where  $A=\left(\begin{array}{cc} a&1\\1&a\end{array}\right)\in PGL_2(\Bbb F_{q})$ is of order $2$ and $U_{q+1}=\{x\in\mathbb{F}_{{q}^{2}}|x^{q+1}=1\}$.
Let $L=(L_1,\ldots, L_{\tau})$ be a support, where $L_1, \ldots, L_{\tau}$ are   some disjoint orbits of $U_{q+1}$ and $|L_i|=2, i=1,\ldots, \tau$.  Let $g(x)=\prod^{s-1}\limits_{j=0}\prod^{1}\limits_{i=0}(x-A^i(\beta^{q^j}))$ be the Goppa polynomial of degree $r=2s$ over $\Bbb F_{q^2}$, and $s$ be the least positive integer such that $A(\beta)=\beta^{q^s}$. Then $\mathcal{A}_{r+1}(v_{g,L},L)$ is a binary irreducible quasi-cyclic parity-check subcode of Goppa code over $\mathbb{F}_{{q}^{2}}$.
\end{cor}
\begin{proof}
By the proof of Theorem $\ref{th3}$,  $g(x)$ is an irreducible polynomial  over $\Bbb F_{q^2}$  and  $\gamma g(x)=(x+d)^rg({A}(x))$, where $0\ne \gamma\in \Bbb F_q$.
  For $x\in U_{q+1}$, ${(A(x))}^{q+1}={(\frac{ax+1}{x+a})}^{q+1}=1$ and $A(x)\in U_{q+1}$. Hence $U_{q+1}$ is globally invariant by the
projective linear map $x \rightarrow A(x)=\frac{ax+1}{x+a}$.  In summary, $\mathcal{A}_{r+1}(v_{g,L},L)$ is a binary irreducible quasi-cyclic parity-check subcode of Goppa code over $\mathbb{F}_{{q}^{2}}$.
\end{proof}
\begin{cor}
Let $\langle A \rangle $ be the cyclic group acting  on $U_n$, where $A=\left(\begin{array}{cc} 0&1\\1&0\end{array}\right)\in PGL_2(\Bbb F_{q})$ is of order $2$ and $U_n=\{x\in\Bbb F_{q^2}| x^n=1, n|(q^2-1)\}$.
Let $L=(L_1,\ldots, L_{\tau})$ be a support, where $L_1, \ldots, L_{\tau}$ are   some disjoint orbits of $U_{n}$ and $|L_i|=2, i=1,\ldots, \tau$.  Let $g(x)=\prod^{s-1}\limits_{j=0}\prod^{1}\limits_{i=0}(x-A^i(\beta^{q^j}))$ be the Goppa polynomial of degree $r=2s$ over $\Bbb F_{q^2}$,  and $s$  be the least positive integer such that $A(\beta)=\beta^{q^s}$. Then $\mathcal{A}_{r+1}(v_{g,L},L)$ is a binary irreducible quasi-cyclic parity-check subcode of Goppa code over $\mathbb{F}_{{q}^{2}}$.

\end{cor}
\begin{proof}
The proof is similar to Corollary $\ref{c12}$. So we omit it.
\end{proof}
All  examples are obtained using the MAGMA system.
\begin{exa}
 Let $\xi$ be a root of an irreducible polynomial $x^5+x^2+1$ over $\Bbb F_2$ and $\Bbb F_{32}=\Bbb F_2(\xi)$. Let $A=\left(\begin{array}{cc} \xi&1\\1&\xi\end{array}\right)\in PGL_2(\mathbb{F}_{32})$. Let $\omega$ be a  primitive element of $\Bbb F_{q^2}$, where $q=32$. The orbits of $U_{q+1}$  are: $L_0=(1)$; $L_1=(\omega^{31}, \omega^{837})$; $L_2=(\omega^{62}, \omega^{558})$;
$L_3=(\omega^{93}, \omega^{806})$; $L_4=(\omega^{124}, \omega^{868}), L_5=(\omega^{155}, \omega^{899}); L_6=(\omega^{186}, \omega^{992}); $ $ L_7=(\omega^{217}, \omega^{930}); L_8=(\omega^{248}, \omega^{341}); L_9=(\omega^{279}, \omega^{651}); L_{10}=(\omega^{310}, \omega^{496}); L_{11}=(\omega^{372}, $ $\omega^{744}); L_{12}=(\omega^{403}, \omega^{434}); L_{13}=(\omega^{465}, \omega^{961}); L_{14}=(\omega^{527}, \omega^{713}); L_{15}=(\omega^{589}, \omega^{620}); L_{16}$ $=(\omega^{682}, \omega^{775})$.

  Let $g(x)=x^2 +\omega^{459}x + \omega^{321}$  be the Goppa polynomial of  Alternant code $\mathcal{A}_{3}(v_{g,L},L)$ over $\mathbb{F}_{{q}^{2}}$. If $L=(L_1,\ldots,L_{16})$, then $\mathcal{A}_{3}(v_{g,L},L)$ is a binary irreducible $16$-quasi-cyclic parity-check subcode of Goppa code over $\mathbb{F}_{{q}^{2}}$ with length $m=32$.
\end{exa}
\begin{exa}

Let $\omega$ be a  primitive element of $\Bbb F_{q^2}$, where $q=32$. Let ${A}=\left(\begin{array}{cc} 0&1\\1&0\end{array}\right)\in PGL_2(\mathbb{F}_{32})$.  The orbits of $U_{31}$  are: $L_0=(1)$; $L_1=(\omega^{33}, \omega^{990})$; $L_2=(\omega^{66}, \omega^{957})$; $L_3=(\omega^{99}, \omega^{924})$; $L_4=(\omega^{132}, \omega^{891})$; $L_5=(\omega^{165}, \omega^{858})$; $L_6=(\omega^{198}, \omega^{825})$; $L_6=(\omega^{198}, \omega^{792})$; $L_7=(\omega^{231}, \omega^{759})$; $L_8=(\omega^{264}, \omega^{726})$; $L_9=(\omega^{297}, \omega^{693})$; $L_{10}=(\omega^{330}, \omega^{660})$; $L_{11}=(\omega^{363}, \omega^{627})$; $L_{12}=(\omega^{396}, \omega^{594})$; $L_{13}=(\omega^{429}, \omega^{594})$; $L_{14}=(\omega^{462}, \omega^{561})$; $L_{15}=(\omega^{495}, \omega^{528})$.

  Let $g(x)=x^2 + \omega^{800}x + 1$  be the Goppa polynomial of  Alternant code $\mathcal{A}_{3}(v_{g,L},L)$ over $\mathbb{F}_{{q}^{2}}$. If $L=(L_1,\ldots,L_{15})$, then $\mathcal{A}_{3}(v_{g,L},L)$ is a binary irreducible $15$-quasi-cyclic parity-check subcode of Goppa code over $\mathbb{F}_{{q}^{2}}$ with length $m=30$.
\end{exa}
\section{Conclusion}

In this paper,  we presented a sufficient and necessary condition for an irreducible monic polynomial $g(x)$ of degree $r$ over $\mathbb{F}_{q}$ satisfying $\gamma g(x)=(x+d)^rg({A}(x))$, where $q=2^n$, $A=\left(\begin{array}{cc} a&b\\1&d\end{array}\right)\in PGL_2(\Bbb F_{q})$, $\mathrm{ord}(A)$ is a prime, $g(a)\ne 0$, and $0\ne \gamma\in \Bbb F_q$.  And we  gave a complete characterization of irreducible polynomials $g(x)$ of degree $2s$ or $3s$ as above, where $s$ is a positive integer. Moreover, we constructed some binary irreducible quasi-cyclic parity-check subcodes of Goppa codes and extended Goppa codes.

\end{document}